**Resistive switching in ferroelectric BiFeO$_3$ by 1.7 eV change of the Schottky barrier height.**

*S. Farokhipoor and B. Noheda\**

Zernike Institute for Advanced Materials, University of Groningen, 9747 AG, The Netherlands.
E-mail: s.farokhipoor@rug.nl, b.noheda@rug.nl

Keywords: ferroelectric resistive switching, BiFeO$_3$ thin films, Schottky barrier tuning, domains and domain walls.

Using metal-ferroelectric junctions as switchable diodes was proposed several decades ago.[1] This was shown to actually work in PbZr$_{1-x}$Ti$_x$O$_3$ (PZT) by Blom et al.,[2] who reported switching in the rectification direction and changes of the current of about 2 orders of magnitude upon switching the polarization direction of the ferroelectric layer. This form of resistive switching enables the read out of a ferroelectric memory state at higher speed compared to the capacitive design, without destroying the information in each reading cycle.[3-5] Recently, Jiang and coworkers have shown that these Schottky barrier effects are enormous in BiFeO$_3$, giving thousand times more switched charge than found by in PZT.[6] Here, by performing local conductivity measurements, we attribute this to a large change of the Schottky barrier height between the as-grown, down-polarized domains and the up-polarized domains. These measurements allow to estimate the relative effect of polarization charges and screening charges on the conduction through the ferroelectric.

Apart from the examples mentioned above and few others,[7] transport through ferroelectrics had not been investigated in detail until less than a decade ago. The reason for that is the large turn-on voltages that were expected for the Schottky behavior of these large band-gap semiconductors.[8] However, as the thin film deposition techniques improve and better layers with lower thicknesses are achieved, metal-ferroelectric-metal heterostructures become more interesting: indeed, the surface charge in a ferroelectric semiconductor plays the same role as the doping charge and it is, in principle, possible to modify the barrier characteristics from Schottky to ohmic without having to dope the material.[2,8]

Resistive switching with ferroelectric layers has attracted a lot of attention in recent years.[6,9-15] One can consider the metal-ferroelectric junction as a Schottky contact, with the ferroelectric polarization modeled as a charged sheet at a certain distance of the physical interface with the electrode,[16,17] in order to account for the so-called 'dead layer'.[18] Assuming small (and non-overlapping) depletion layers, this model has given experimental Schottky barrier height (SBH) values in the range of $\Delta\phi$ = 0.2-0.9eV for various ferroelectrics and orientations.[19-21] However, in order to model realistic metal-ferroelectric-metal heterostructures, considering the incomplete screening of the polarization surface charges is needed.[22] For 'bad' metal electrodes (large screening lengths) and for ultrathin ferroelectric layers as those requested for tunnel-junctions[23,24] the charge at the electrode may not be enough to screen the polarization and a large depolarization field may be present.[25,26]

Classically, the origin of the 'dead layer' (non-switchable interface layer) was assumed to be associated to defects in the film. But even in the case of an ideal, defect-free, ferroelectric layer, a certain 'dead-layer' or series capacitance always exists due to different ionic displacements at the interface, which could be in some instances more important than the electronic effects due to imperfect screening[27-29] or even give rise to a super-polarized layer ('negative dead layer').[30] It is worth noticing that in all cases (defective interfacial layers, Thomas-Fermi screening lengths or different ionic displacement at the interface), the system can be modeled by adding an interfacial capacitance in series with the ferroelectric capacitance.[27-30]

Following the Thomas-Fermi approach, Zhuralev et al.[31] pointed out that using electrodes with different screening lengths would lead to changes of several orders of magnitude in the tunneling conductance through a ferroelectric upon polarization switching (*giant electroresistance effect*).[22,31] Less discussed has been the way in which ferroelectric switching modifies the resistance through thicker layers (when direct tunneling is not possible) in case of incomplete screening at the electrodes. In this case of thicker layers, the average potential barrier height across the ferroelectric layer is not the determining parameter for conduction and resistive switching can be obtained with symmetric electrodes.[27]

First-principles calculations that consider the ferroelectric as a perfect insulator, predict changes in the SBHs of up to 2.6eV upon ferroelectric switching, in the case of $PbTiO_3$ (large ferroelectric polarization) with $SrRuO_3$ electrodes (large effective screening length).[27] A 1D electrostatic model, that takes into account conduction through the ferroelectric as well as the imperfect screening at the electrodes, has recently shown that the SBH changes induced at the metal- $BiFeO_3$ interface are significantly larger at the Pt electrode than at the $SrRuO_3$ electrode.[12,32] This is in contradiction with first principles calculations[25,27,30] and recent experiments[33] that show smaller SBH changes for the noble metal electrodes (with smaller effective screening lengths). Moreover, it is believed that the presence of vacancies and charged defects plays an important role determining conduction through the ferroelectric layers (in fact, resistive switching is often found in non-ferroelectric materials)[34] so that both polarization charges and defect charges need to be considered to understand the resistive switching phenomena in ferroelectrics.[6, 10-15] However, there does not seem to be a clear understanding about the relative importance of the two.

In order to shed light on the open problems discussed above, we have locally measured the SBHs in down-polarized domains, up-polarized domains and non-polar domain walls (DWs) of epitaxial BiFeO$_3$, a ferroelectric that is known to display remarkable resistive switching.[6,11-12] We measure a change of the SBH as large as 1.7eV between the as-grown down-polarized domains and the up-polarized domains, when we use a Cr-coated tip of an AFM microscope as the top electrode and an epitaxial SrRuO$_3$ layer as bottom electrode. These measurements allow to estimate the relative effect of polarization charges and screening charges on the conduction through the ferroelectric.

BiFeO$_3$ thin films with thicknesses of 70 nm were grown by pulsed laser deposition on single-terminated (001)-oriented SrTiO$_3$ substrates covered by a 5nm thick buffer electrode layer of SrRuO$_3$ (see also experimental section).[35,36] The as-grown films were self-polarized with the polarization pointing down (towards the substrate), that is, with only four of the eight pseudo-rhombohedral domains present.[37] This preferential poling is common in ferroelectrics[38] and could be due to deviations in the surface chemistry or be caused by the preferential termination during growth.

In BiFeO$_3$ thin films similar to those used here (same substrate, same growth conditions, same lab), enhanced conduction through the DWs with respect to that of the domains has been shown to be determined by the height of the Schottky barrier at the metal-semiconductor interface; the SBH is shown to be significantly reduced at the domain walls.[35,36] That is consistent with earlier reports by Pintilie et al. showing that the SBH determines the current through a thick BiFeO$_3$ films when using macroscopic electrodes.[19] **Figure S1** shows the apparent SBH as a function of $V^{1/2}$, from the extrapolation to the ordinate axis, the (zero-voltage) SBHs are obtained.[19] The

SBH was measured to be as high as $\phi_\downarrow = 2.6(6)$ eV in the as-grown, down-polarized domains.[35] All the experiments were consistent with a n-doped BiFeO$_3$ layer (as expected from the presence of oxygen vacancies) and injection of electrons from the top Cr-electrode.

To a first approximation, the difference between the work function of the metal and the electron affinity of BiFeO$_3$ gives rise to the SBH[39,40] of about $\phi_o = 1.4$ eV for the top Cr-BiFeO$_3$ interface,[41] if we considered non-polar BiFeO$_3$. A polarization pointing down inside the ferroelectric domains will induce negative surface charges at the top interface and will increase the SBH for electrons injected from the Cr electrode, as indeed observed.

In order to further investigate the effect of polarization, we locally switch the out-of-plane component of the polarization up, such that now positive bound charges are induced at the top interface. This should decrease the barrier for electronic injection. Indeed, as observed by other groups in bulk and thin film samples,[6, 11-12] a much larger conduction level is observed in the up-polarized regions (see **Figure 1**). The I-V curves measured at different temperatures (at and above room temperature) both at the up- and down-polarized domains are shown in **Figure S2**. In **Figure 2** the up-polarized Schottky-Richardson-Simmons[42] plots are shown (for those corresponding to the down-polarized domains see refs.[35,36]). From these curves, the apparent Schottky barrier heights as a function of $V^{1/2}$ can be obtained (see **Figure 2b**) and a zero-voltage SBH of $\phi_\uparrow = 0.9(1)$ eV can be extracted from the intercept at the origin.[19] Therefore, polarization switching induces a change of 1.7eV in the SBH, which explains the huge change in the conduction level observed by us and various authors upon switching the polarization of a BiFeO$_3$ crystal.[6, 11-12]

Even though the polarization bound charges at the interface of non-perfect electrodes seem to be the crucial ingredients to explain this change of SBH (and thus conduction through the ferroelectric),[27,32] it is oftern reported that oxygen vacancies play an important role in resistive switching. In order to separate the effect of polarization charges from that of charged defects it is useful to turn to the local measurements at DWs because these are non-polar entities within the ferroelectric. When measuring locally at non-polar 71° DWs, SBHs of $\phi_{DW}$ = 0.8(2)eV and $\phi_{DW}$ = 0.5(6)eV have been observed in samples with lower and higher oxygen content, respectively,[35,36] as shown in **Figure S1**. This indicates that the oxygen vacancies in the two samples lower the SBH by $\Delta\phi_V = \phi_o$ - 0.8eV = 0.6eV and $\Delta\phi_V = \phi_o$ - 0.5eV = 0.9eV, respectively. When comparing the conduction levels for samples with different oxygen content, we observe that while the current at the domain walls could be increased by a factor of 10 simply by increasing the cooling rate after growth (therefore increasing the number of vacancies), the current of the domains remained unchanged (see **Figure 3**). This seems to confirm the model of positively charged oxygen vacancies accumulating around the walls and reducing the SBH with the electrode.[35,36]

Thus, from the results of **Figure 3**, we can then assume that the oxygen vacancies do not contribute significantly to lowering the SBH in the domains. In the as-grown down domains, the SBH with the top electrode is then modified mainly by the negative polar charges, as $\phi_\downarrow = \Delta\varphi_o + \Delta\varphi_P$ = 1.4eV + $\Delta\phi_P$ = 2.6eV (measured value) and, thus the effect of polarization bound charges is $\Delta\phi_P$ = 1.2eV. Upon polarization switching, the SBH will then become $\phi_\uparrow = \phi_o - \Delta\phi_P + \Delta\phi_{scr}$ = 1.4eV - 1.2eV + $\Delta\varphi_{scr}$ = 0.9eV (measured value), which indicates the presence of a different screening mechanism contributing $\Delta\phi_{scr}$ = 0.7eV to the total SBH of the up-domains. The high conduction state in the up-polarized domains is stable at room temperature and can be measured for weeks after poling. However, increasing the temperature to (and above) 180°C the up-polarized regions switch back within one hour and the low conduction state is recovered (see **Figure S3**). This points to adsorbates as screening charges of the (artificially polarized in air) up-domains. If the screening adsorbates are removed at high temperatures, the depolarization field makes the up-polar domains unstable.

First-principles calculations indicate that the largest changes in the SBH upon polarization switching are to be expected at the interface with the 'worse' of the two electrodes.[27] In $BiFeO_3$ films with bottom epitaxial $SrRuO_3$ layer and top Cr or Pt electrodes, the experiments indicate that the largest change takes place at the top electrode, despite the fact that is a 'better' metal. The reason for that is probably the additional defective 'dead-layer' induced at the top interface because of the chemical and crystallographic mismatch between the metals and the film structures. This additional dead-layer at the top interface or extra series-capacitance in the problem could be mimicked by considering an artificially positive dielectric constant at the metal electrodes.[32]

One can expect that diffusion of oxygen vacancies and other charged defects inside the ferroelectric is easier in poly-crystalline samples or films showing columnar growth than in fully epitaxial films. Therefore, the exact interplay between the polarization and defect charges varies considerably with the sample morphology explaining the different behaviors reported in the literature.[6, 15] Moreover, since this is mainly an interface effect, the type of structural interfaces and twinning is of crucial importance to determine the dominant type of conduction. The films discussed here exhibit only in-plane twinning, that is, the (001) planes of the film are parallel to those of the substrate,[37] which should be favorable for conduction.[19,37] It is also worth to mention that the in-plane twinning gives rise to a more clamped structure with larger coercive fields than when the interfaces are buckled by out-of-plane twinning.[19,43] In fact, we can apply up to 5 V without switching the polarization or inducing domain movement, which allows to increase the current values up to 15 pA at room temperature. Given that the tip diameter, analyzed by Scanning Electron Microscopy, is in between 50nm (new tip) and 100nm (used tip), the observed current densities can be as high as 0.5 A/cm$^2$ at room temperature.

In summary, we have measured a 1.7eV change in the SBH when switching the polarization of a 70 nm thick BiFeO$_3$ film from (as-grown) down polarized to up polarized. This explains the huge changes in resistance for bias voltage applied parallel or antiparallel to the polarization direction, as previously observed. The SBH values measured locally at up-domains, down-domains and non-polar domain walls allow us to separate the polarization contributions from the screening charge contributions to the SBH. Current densities in the forward direction of 0.5 A/cm$^2$ are measured. Even larger current values of 5.4 A/cm$^2$ have been obtained in different BiFeO$_3$ films.[6] These values rival the dendritic short mechanisms investigated in non-ferroelectric resisitive switching.[34]

**Experimental Section**

High quality epitaxial BiFeO$_3$ thin films with thicknesses of 70 nm were grown by pulsed laser deposition on single-terminated (001)-oriented SrTiO$_3$ substrates covered by a 5nm thick buffer electrode layer of SrRuO$_3$, as described elsewhere.[35,36] After growth, the oxygen pressure in the growth chamber was increased to 100 mbar and the films were cooled down at a rate of 3$^o$C/min.

Atomic force microscopy (AFM), as well as piezo-force microscopy (PFM) and conductive AFM (c-AFM) were carried out using Nanoscope Dimension V microscope fromVEECO (now

Bruker). For the c-AFM measurements a TUNA$^{TM}$ amplifier (VEECO/Bruker) was used. The TUNA$^{TM}$ module allows three different amplification gains with current sensitivities ranging from 0.5 pA to 24 pA. The tip was grounded and the conductivity was mapped by applying a DC bias to the bottom electrode. Cr/Co-coated n-doped Si AFM tips were used as top electrodes and the SrRuO$_3$ buffer layer as bottom electrode.

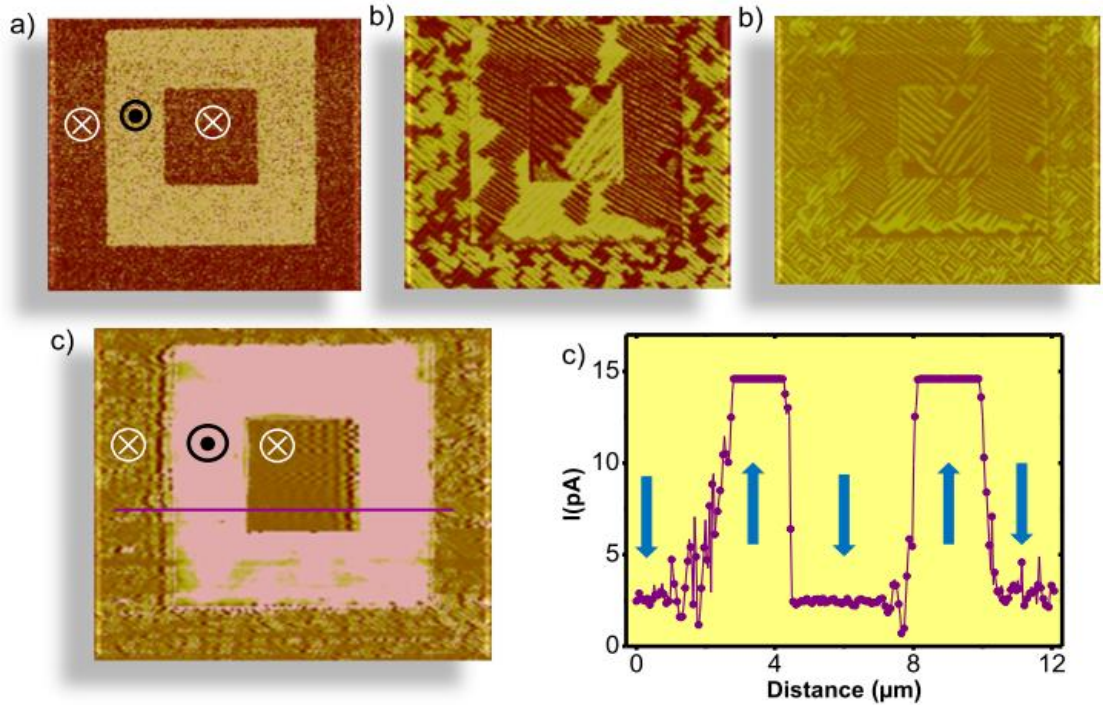

**Figure 1.** (a) Out-of-plane piezo-force microscopy (PFM) phase image showing the as-grown down-polarized state (cross, dark contrast). A square of polarization up (dot, light contrast) has been written. In a second inner square the polarization has been switched back down. (b) In-plane phase and amplitude PFM images, use to infer the types and configuration of domain walls. (c) conductive-AFM (C-AFM) image of the same area of the film showing enhanced currents at the up-polarized area compared to those in the as-grown state and down polarized regions. The applied voltage is 2.9V. All areas are 20μm x20μm (c) Line scan along the line depicted in (c) across regions with different polarizations states, showing the magnitude of the current. The TUNA amplifier is saturated in the up-polarized domains.

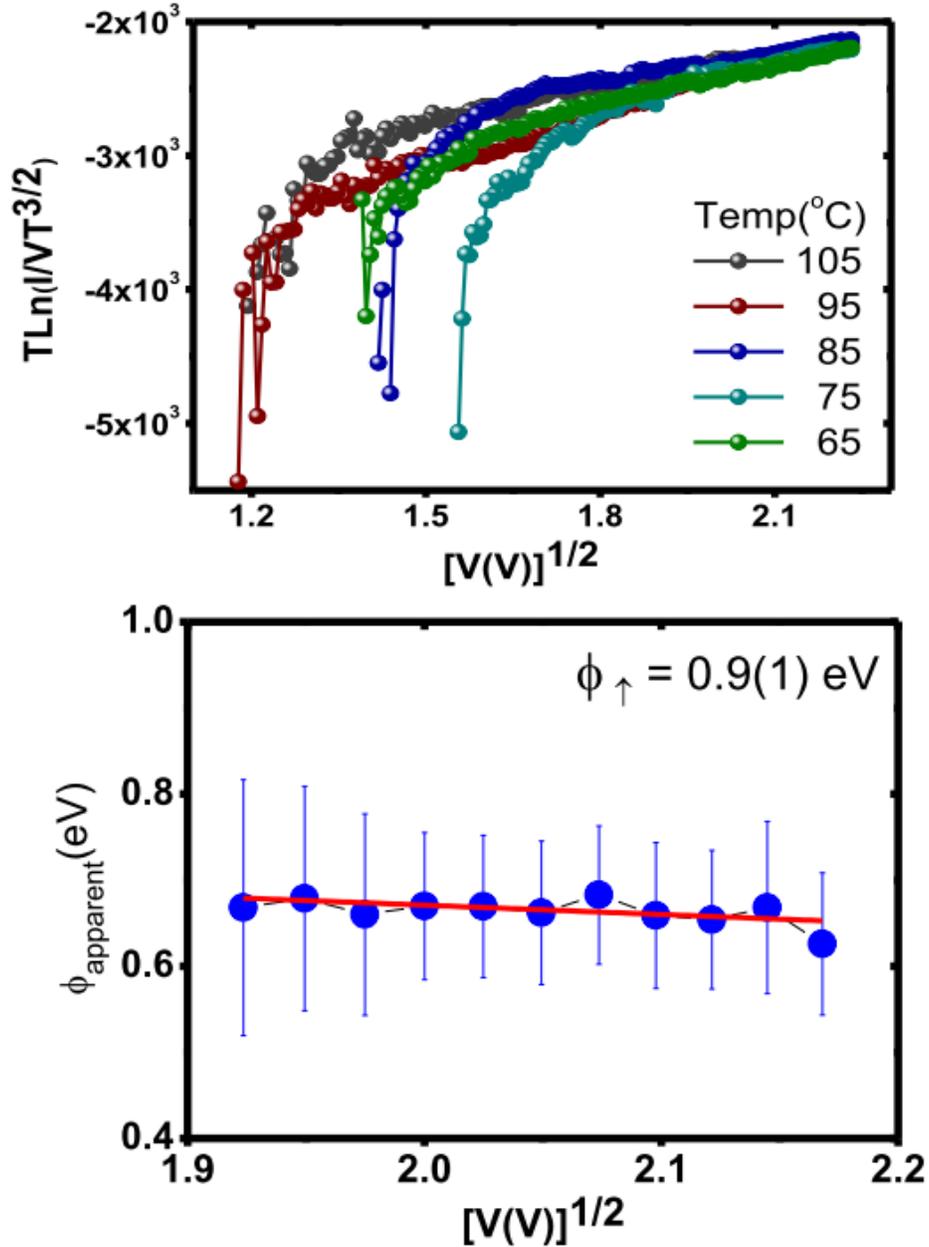

**Figure 2.** The IV curves corresponding to artificially switched up-domains are plotted in such a way that those data points that fulfill the Simmons-Richardson-Schottky equation[42] show a linear dependence and collapse in a single line. From the slope of the curves a dielectric permittivity of 7.6(3) is obtained (top). Apparent SBH as a function of $V^{1/2}$ (bottom). From its ordinate at V= 0, a SBH of 0.9(1)eV is obtained.

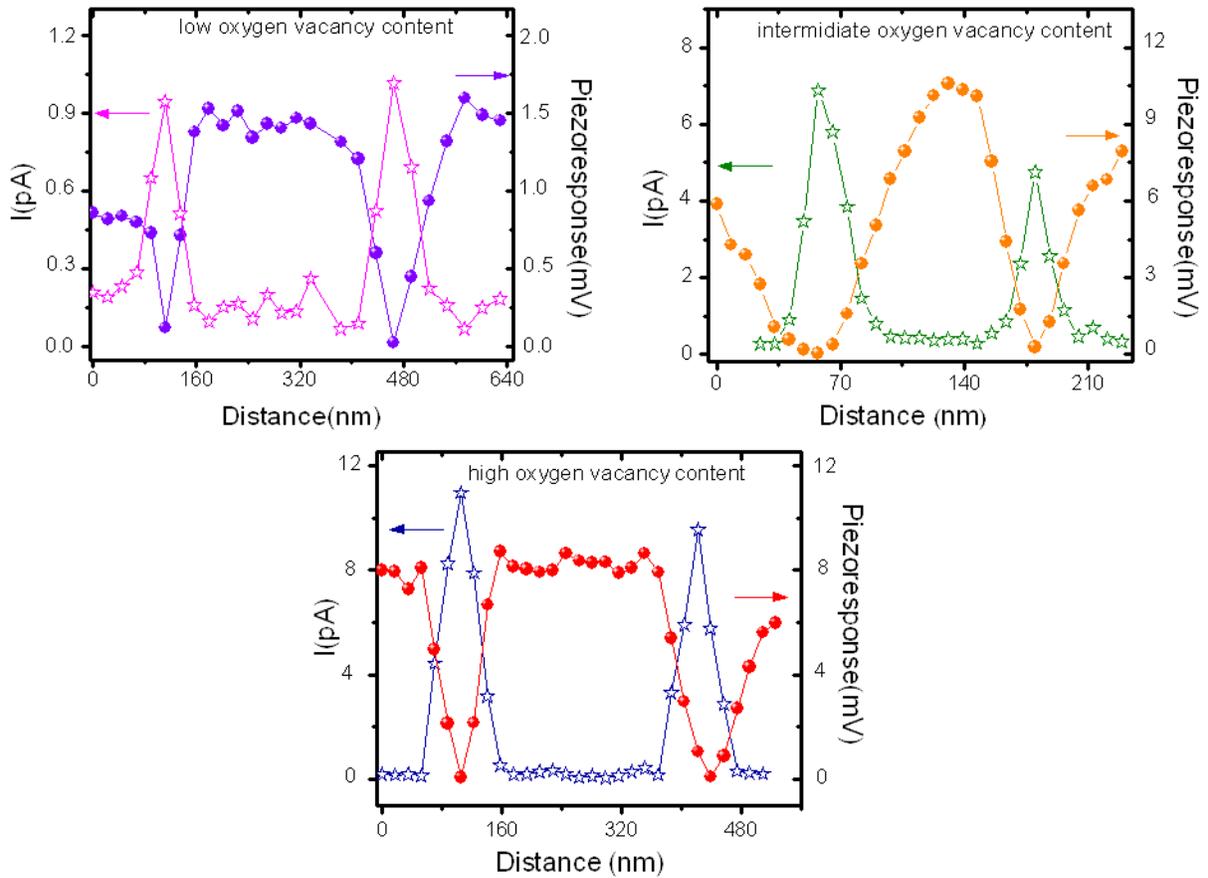

**Figure 3**. Line scan of both piezoresponse amplitude (shown by stars) and the current (shown by spheres) across one domain with two adjacent domain walls of three samples grown under identical conditions but with different annealing conditions; (a) annealing pressure of $PO_2$= 300 mbar and a cooling rate of 1 $^{o}$C/min, (b) $PO_2$=100mbar and rate of 3 $^{o}$C/min and (c) $PO_2$=100mbar and rate of 40 $^{o}$C /min.

# Supplementary information

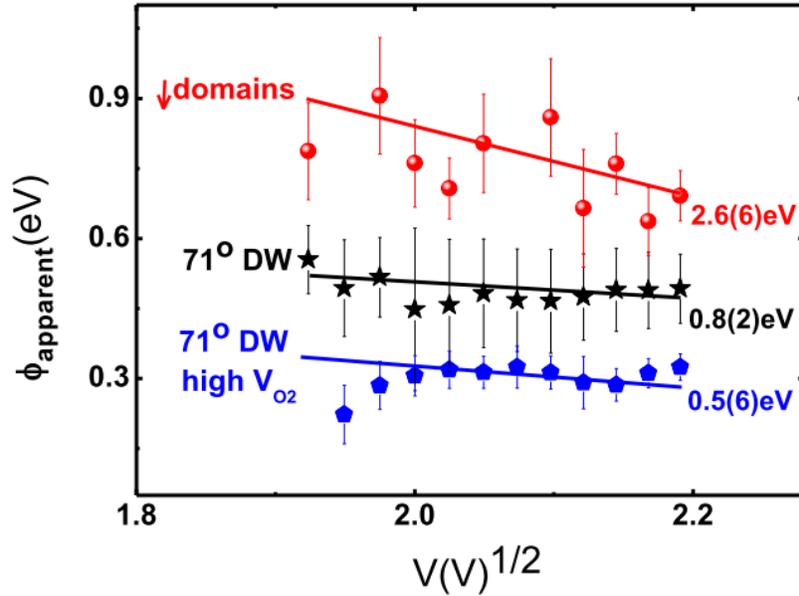

**Figure S1.** The so-called *apparent* Schottky barrier height (SBH) is defined as $\phi_{apparent} = \phi_o - e\sqrt{\frac{eV}{4\pi\varepsilon_o\varepsilon_\infty d}}$, where e is the electron charge, V is the applied bias, $\varepsilon_o$ is the dielectric permittivity of vacuum and $\varepsilon_\infty$ is the optical dielectric permittivity of $BiFeO_3$. When plotted as a function of $V^{1/2}$, the intercept at the origin shows the experimental (zero voltage) SBH at the down-domains (red circles) and at $71^o$ domain walls of two samples with less (black stars) and more (blue pentagons) oxygen vacancies. These are plotted from data of references [35,36]

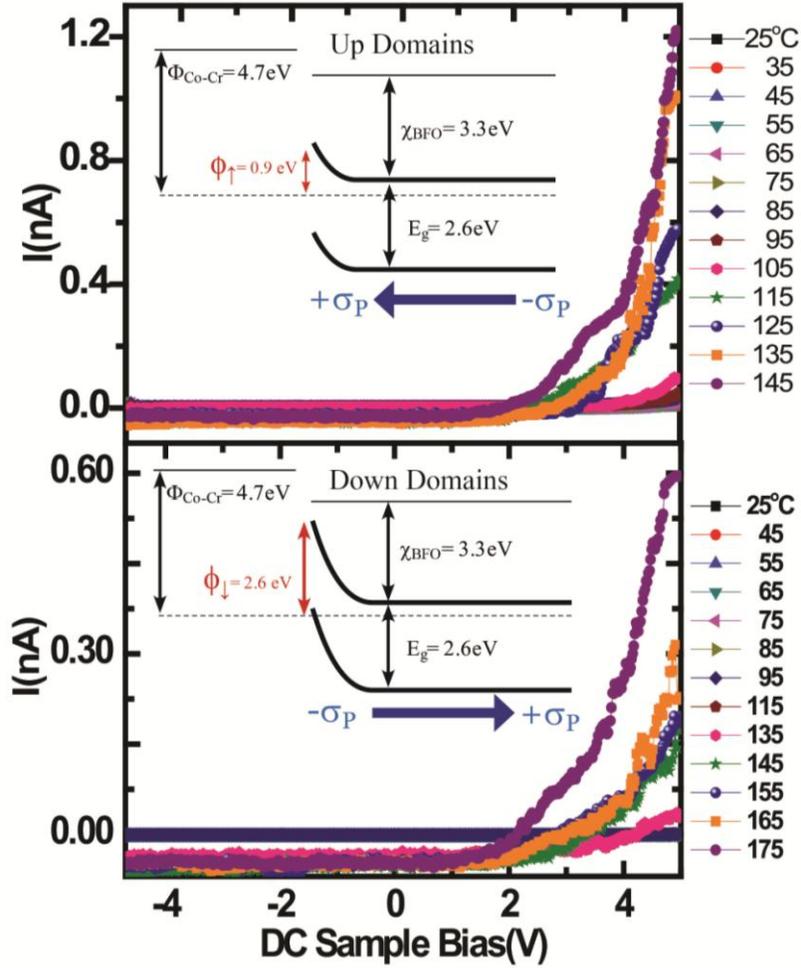

**Figure S2.** I-V curves measured at different temperatures in the up-domains (a) are compared to those previously measured in the down-domains[33] (b). The insets show sketches of the electronic band structure at the interface in both cases.

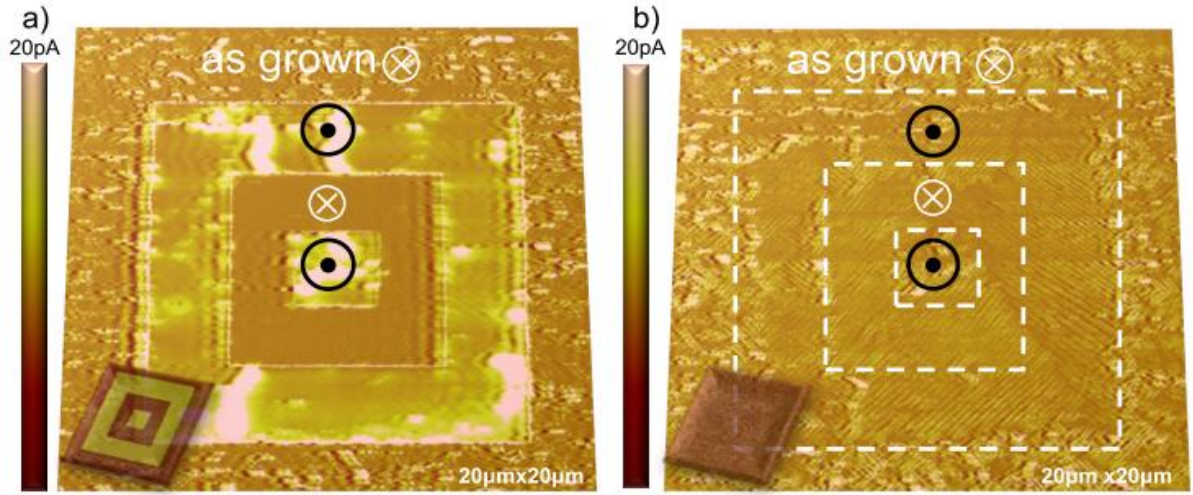

**Figure S3.** Conduction map of the sample surface after writing an up-polarized square, an inner down-polarized one and still a smaller up-polarized square in the center (a). The encircled dots and crosses denote the regions with polarization up and down, respectively. The imaging voltage used was 2.4V, smaller than that of Figure 2 in order to avoid saturation of the TUNA amplifier. The out-of-phase PFM image is also shown as an inset. (b) shows a second conduction map under the same DC voltage bias after annealing the sample for one hour at 200°C. As it can be seen, there is no contrast in the out-of-plane PFM inset, indicating that the polarization of the written areas has switched back. Correspondingly, the current has decreased and only current through the domain walls is observed.